\documentclass[aps,prl,twocolumn,showpacs,superscriptaddress,groupedaddress]{revtex4-1}  
\usepackage{graphicx}  
\usepackage{dcolumn}   
\usepackage{bm}        
\usepackage{amssymb}   
\usepackage{color}

\begin{document}


\title{Hydrodynamic theory of freezing: Nucleation and polycrystalline growth}

\author{Frigyes Podmaniczky,$^1$ Gyula I. T\'oth,$^{1,2}$ Gy\"orgy Tegze,$^1$ and L\'aszl\'o Gr\'an\'asy$^{1,3}$}

\affiliation{$^1$Research Institute for Solid State Physics and Optics,
P.O. Box 49, H-1525 Budapest, Hungary}

\affiliation{$^2$Department of Physics, University of Bergen, 
All\'egaten 55, 7005 Bergen, Norway}

\affiliation{$^3$BCAST, Brunel University,
Uxbridge, Middlesex, UB8 3PH, United Kingdom}

\date{\today}

\begin{abstract}
Structural aspects of crystal nucleation in undercooled liquids are explored using a nonlinear hydrodynamic theory of crystallization proposed recently [G. I. T\'oth {\it et al.}, J. Phys.: Condens. Matter {\bf 26}, 055001 (2014)], which is based on combining fluctuating hydrodynamics with the phase-field crystal theory. We show that in this hydrodynamic approach not only homogeneous and heterogeneous nucleation processes are accessible, but also growth front nucleation, which leads to the formation of new (differently oriented) grains at the solid-liquid front in highly undercooled systems. Formation of dislocations at the solid-liquid interface and interference of density waves ahead of the crystallization front are responsible for the appearance of the new orientations at the growth front that lead to spherulite-like nanostructures.\end{abstract}

\maketitle
Crystal nucleation, i.e., stochastic formation of crystal grains via fluctuations that are able to grow plays an essential role in the making of polycrystalline and nanostructured matter \cite{ref1, ref2}, including the formation of {\it polycrystalline growth forms} such as disordered dendrites \cite{ref3}, and spherulites \cite{ref4}. The latter structures appear via the formation of new grains at the solidification front, a process termed “growth front nucleation” (GFN \cite{ref5}). This phenomenon has been successfully addressed by conventional phase-field methods relying on coarse grained fields \cite{ref5}. Unfortunately, these works cannot provide details on the micromechanism of GFN. Although molecular simulations and theory recently shed light on many details of nucleation and liquid ordering preceding nucleation \cite{ref6}, GFN seems to be out of scope for such studies. Continuum theories working on the molecular scale offer a complementary approach to molecular  simulations, and may deliver additional information on the relevant nanoscale processes, such as the mechanism of polycrystalline growth.

A fairly successful continuum approach, termed the Phase-Field Crystal (PFC) model, was developed recently to address the microscopic aspects of crystallization \cite{ref7,ref8}. The majority of the PFC studies were done assuming diffusive dynamics, which approximates reasonably crystalline aggregation in suspensions of micron size colloidal particles. In a recent work, we have proposed a hydrodynamic theory of freezing (HPFC) \cite{ref9} that applies for solidification in normal liquids. Our approach relies on fluctuating nonlinear hydrodynamics\cite{ref10}, and employs the free energy functional of the PFC model in determining the reversible stress tensor. This model recovers the proper dispersion relation for long wavelength acoustic phonons, a steady state front velocity, which is inversely proportional to the viscosity (as opposed to the time dependent front velocity observed in the case of diffusive dynamics), and describes the stress relaxation reasonably \cite{ref9}. It is thus expected to be able to capture defect formation, and therefore polycrystalline growth on the nanoscale. 
	
Herein, we first employ the HPFC model for describing homogeneous and heterogeneous crystal nucleation, and then to the formation of new grains at the solidification front (GFN). While there were other PFC-based hydrodynamic models put forward recently \cite{ref11}, it is only the HPFC for which steady state growth, $v \propto \mu_S^{-1}$, and proper capillary wave spectrum were demonstrated.  

In the HPFC model, we start from momentum transport and continuity equations used in fluctuating nonlinear hydrodynamics \cite{ref10}:  
\begin{equation}
\label{eq:NS}
\frac{\partial \mathbf{p}}{\partial t} + \nabla \cdot (\mathbf{p} \otimes \mathbf{v})  = 
\nabla \cdot \bigg [ \mathbf{R}(\rho) + \mathbf{D} (\mathbf{v}) + \mathbf{S} \bigg ], \\
\end{equation}
\begin{equation}
\label{eq:cont}
\frac{\partial \rho }{\partial t} + \nabla \cdot \mathbf{p} = 0.
\end{equation}
Here $\mathbf{p}(\mathbf{r}, t)$ is the momentum, $\rho(\mathbf{r}, t)$ the mass density, $\mathbf{v} = \mathbf{p}/\rho$ the velocity, $\nabla \cdot \mathbf{R} = - \rho \nabla \frac{\delta F[\rho]}{\delta \rho} \approx - \rho_0 \nabla \frac{ \delta F[\rho]}{\delta \rho}$ the divergence of the reversible stress tensor, $\frac{ \delta F[\rho]}{\delta \rho}$ the functional derivative of the free energy with respect to density, $\rho_0$ a reference density, and $\mathbf{D} = \mu_S \{ (\nabla \otimes \mathbf{p}) + (\nabla \otimes \mathbf{p})^T \} + [\mu_B - \frac{2}{3}\mu_S] \mathbf{I} (\nabla \cdot \mathbf{v}) $ the dissipative stress tensor, $\mu_S$ and $\mu_B$ are the shear and bulk viscosities, respectively, while the fluctuation-dissipation theorem yields the following covariance tensor for the momentum noise $\mathbf{S}$:
\begin{eqnarray}
\label{eq:eq7}
\langle S^{\mathbf{r},t}_{ij}S^{\mathbf{r}',t'}_{kl}\rangle &=& \left( 2 k_B T \mu_S \right) \delta(\mathbf{r}-\mathbf{r}') \delta(t-t') \times \\
\nonumber && \times \left[ \delta_{ik}\delta_{jl}+\delta_{jk}\delta_{il} + \left( \frac{\mu_B}{\mu_S} - \frac{2}{3} \right) \delta_{ij}\delta_{kl} \right] \enskip .
\end{eqnarray}
Here $k_B$ is Boltzmann's constant and $T$ the temperature.

We adopt the free energy functional of the PFC model with parameters used in Ref. \cite{ref9}:
\begin{equation}
\label{eq:eq11}
\frac{F}{n_0 k_B T} = \int dV \left\{ f(n) - \frac{C_2}{2}(\nabla n)^2 - \frac{C_4}{2}(\nabla^2 n)^2 \right\} \enskip ,
\end{equation}
where the local free energy density is
\begin{equation}
\label{eq:eq12}
f(n)=(1-C_0)\frac{n^2}{2}- \left(\frac{a}{2}\right)\frac{n^3}{3} + \left(\frac{b}{3} \right)\frac{n^4}{4} \enskip .
\end{equation}
The order parameter is the normalized mass density: $n=\tilde{\rho}-1$, where $\tilde{\rho}=\rho/\rho_0$ ($\rho_0=m_0 n_0$ is the reference density, where $m_0$ is the atomic mass and $n_0$ the number density of the reference liquid). $C_0$ can be related to the bulk modulus of the reference liquid via $K_0=(1-C_0)(n_0 k_B T)$, while $C_2$ and $C_4$ are responsible for elasticity \cite{ref9}. This formalism transforms into the usual Swift-Hohenberg formalism as follows \cite{ref8}:
the reduced temperature and reduced density are expressed as $\epsilon = -(r - t^2/3)$ and $\psi = \phi + t/3$, where $r = 4 (1 - C_0) Z -1$,  $\phi = n/X$, $t = - a \sqrt{3 Z/b}$, $X = \sqrt{3 /4 b Z}$, and $Z = |C_4| /C_2^2$. 
The HPFC model differs from the original PFC model in only the dynamic equations. 

\begin{figure}[t]
\begin{tabular}{ccc}
&(a)\includegraphics[width=3cm]{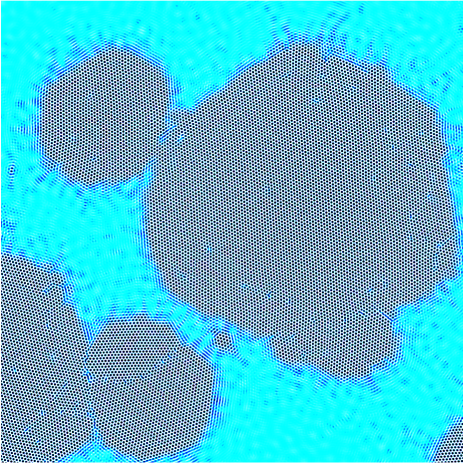} 
&(b)\includegraphics[width=3cm]{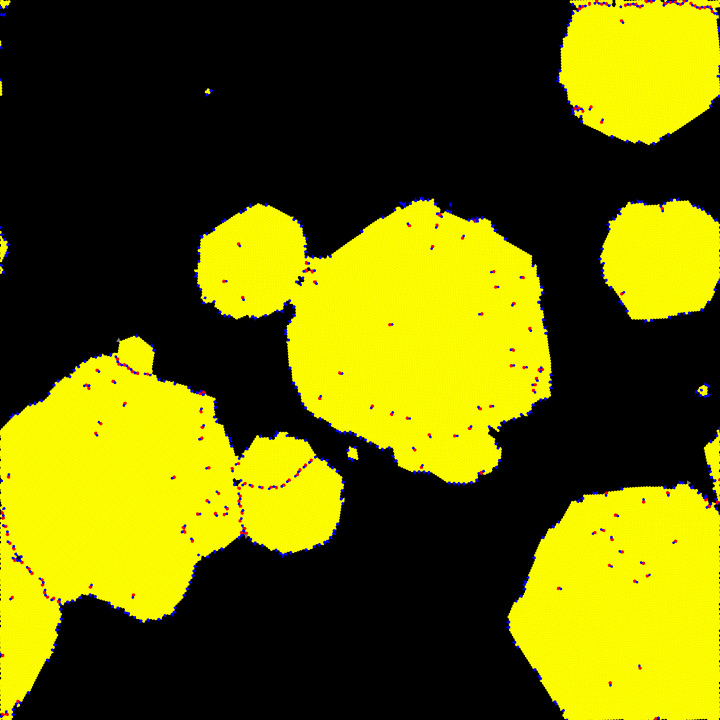}\\ 
&(c)\includegraphics[width=3cm]{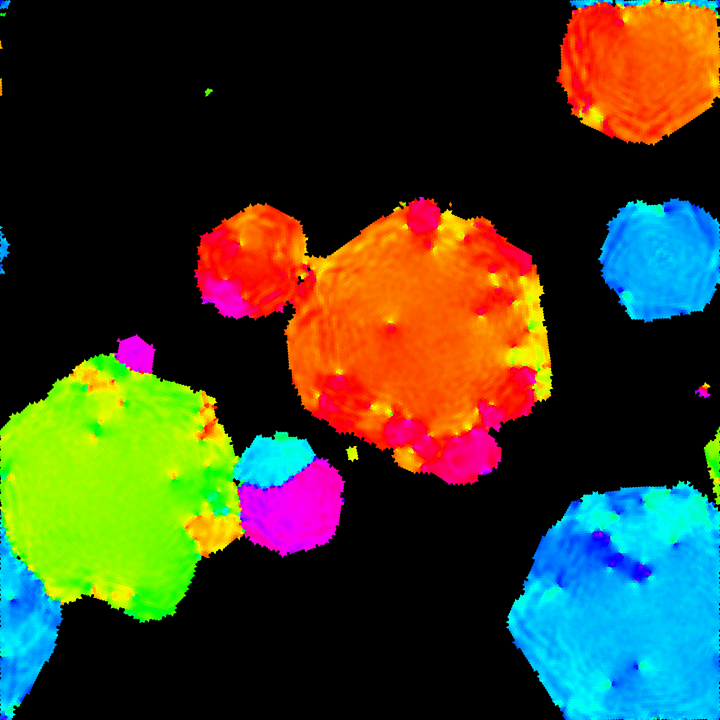} 
&(d)\includegraphics[width=3cm]{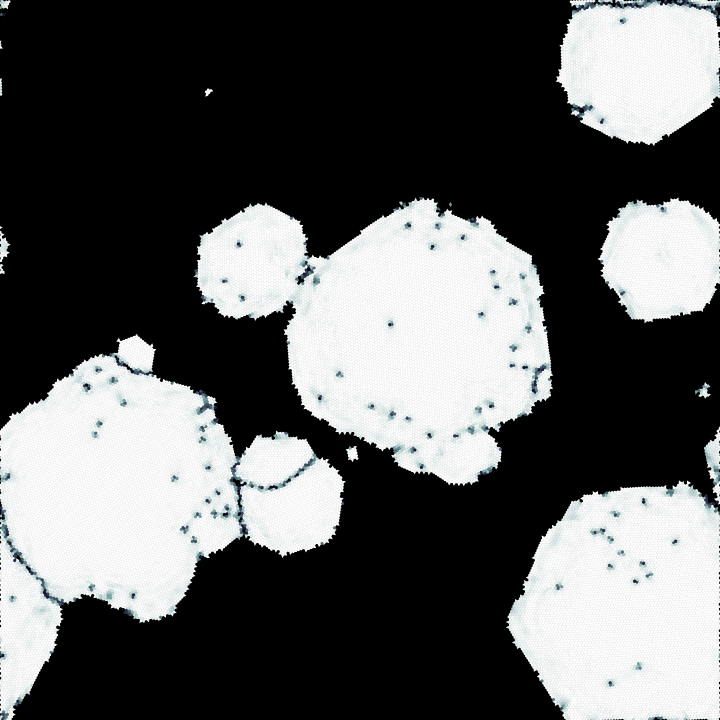}\\ 
&   \includegraphics[width=4cm]{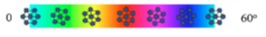}
\end{tabular}
\caption{(color online) Snapshots of density, Voronoi, orientation, and disorder maps for a HPFC simulation performed with noise slightly backward from the liquid stability limit ($\epsilon = 0.1158$, and $\psi = -0. 1982$) on a $2048^2$ square lattice. The local orientations corresponding to coloring in (c) are also shown. The whole simulation box is shown in panels (c)--(d), and the central quarter in (a). The color scale of panel (a) was chosen so that it enhances the visibility of the density waves at the solid-liquid interface. 
}
\label{fig:fields}
\end{figure}

The combination of fluctuating hydrodynamics with atomic scale theory is supported by recent results, which indicate that fluctuating hydrodynamics remains valid down to the nanoscale \cite{ref12}. To avoid interatomic flows in the crystal owing to the large density gradients, we employ coarse-grained momentum and density fields when computing the velocity field: $\mathbf{v} = \hat{\mathbf{p}}/\hat{\rho}$, an approximation used in the advective and viscous dissipation terms \cite{ref9}. While the HPFC model is not restricted to two dimensions (2D), an extensive testing was performed so far in 2D (see Ref. \cite{ref9}). 

Following Ref. \cite{ref9}, the properties of liquid iron were used to fix the model parameters for the 2D simulations. The kinetic equations were solved in 2D, using a pseudo-spectral scheme with a second order Runge--Kutta time stepping, while employing periodic boundary conditions on square grids of sizes ranging from $2048^2$ to $8196^2$. Accordingly, the presented results refer to a hypothetical 2D iron, which can in principle be realized in molecular dynamics simulations, and are expected to be relevant to crystallization in thin metal films. The melting point corresponds to $\epsilon_L = 0.0923$, whereas the scaled liquid density is $\psi_L = - 0.1982$. The linear stability limits taken at constant density or at constant temperature are $\epsilon_c = 0.1178$ and $\psi_c = - 0.1754$. The reduced temperature, and thus the undercooling, was tuned by varying $C_0$.

\begin{figure}[t]
\includegraphics[width=8.5cm]{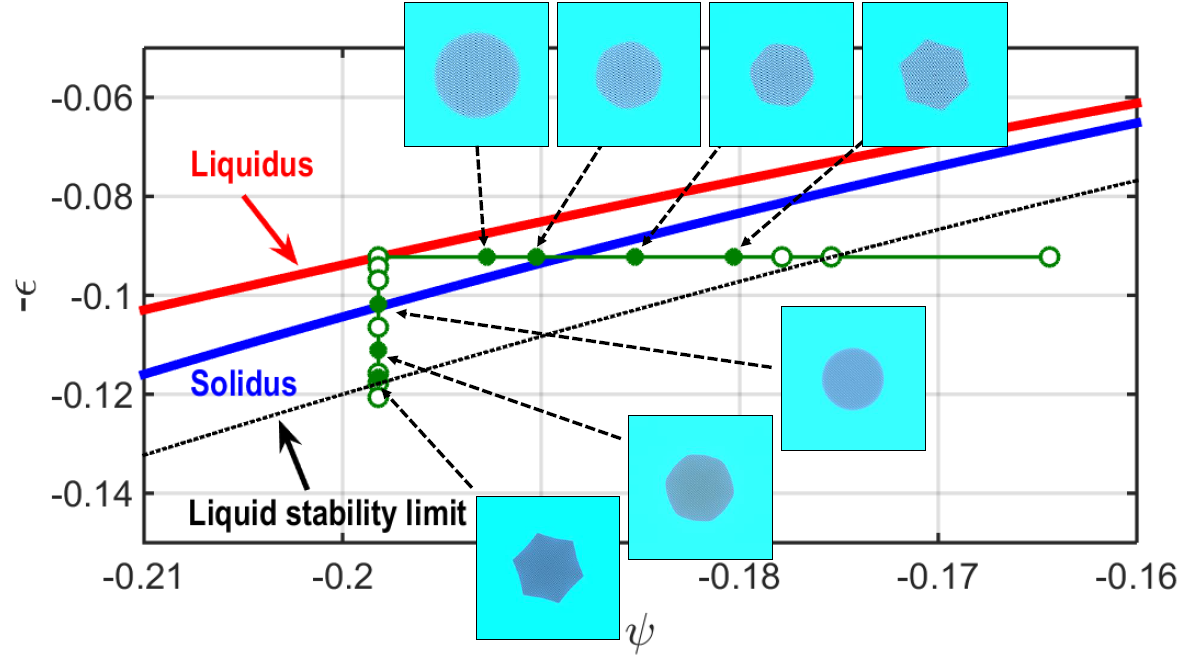}
\caption{(color online) Section of the phase diagram, in which the HPFC simulations  were performed without noise (circles). The heavy red, blue lines and the black dotted lines stand for the liquidus, solidus curves, and the linear stability limit of the liquid. $\epsilon$ is the reduced temperature (distance from the critical point), whereas $\psi$ is the reduced particle density. The blue inserts show the variation of growth shapes (density maps) along the $\epsilon = const.$ and $\psi = const.$ lines.   
}
\label{fig:PHD}
\end{figure}

The structure of the solid phase is characterized by the number density map, whose peak positions are analyzed in terms of Voronoi polygons, and the bond-order parameter, $g_6 = \sum_j exp\{ i 6\theta_j\}$, where $\theta_j$ is the angle corresponding to the $j$-th neighbor in the laboratory frame. The crystal grains in polycrystalline cases were identified on the basis of the orientation map obtained as the phase angle of the complex hexatic bond-order parameter $g_6$. The Voronoi polygons were colored gray, blue, yellow and red, when having 4, 5, 6, and 7 neighbors, respectively. $|g_6|$ characterizes the degree of disorder, and its phase specifies the local crystallographic orientation. Examples of these fields are displayed in Fig. \ref{fig:fields}.

\begin{figure}[t]
\begin{tabular}{ccc}
(a)\includegraphics[width=3.cm]{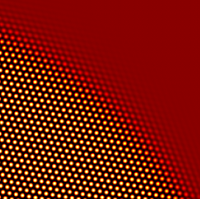} 
(b)\includegraphics[width=3.cm]{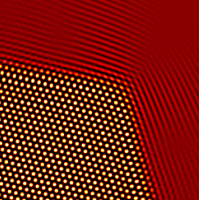}\\ 
(c)\includegraphics[width=7cm]{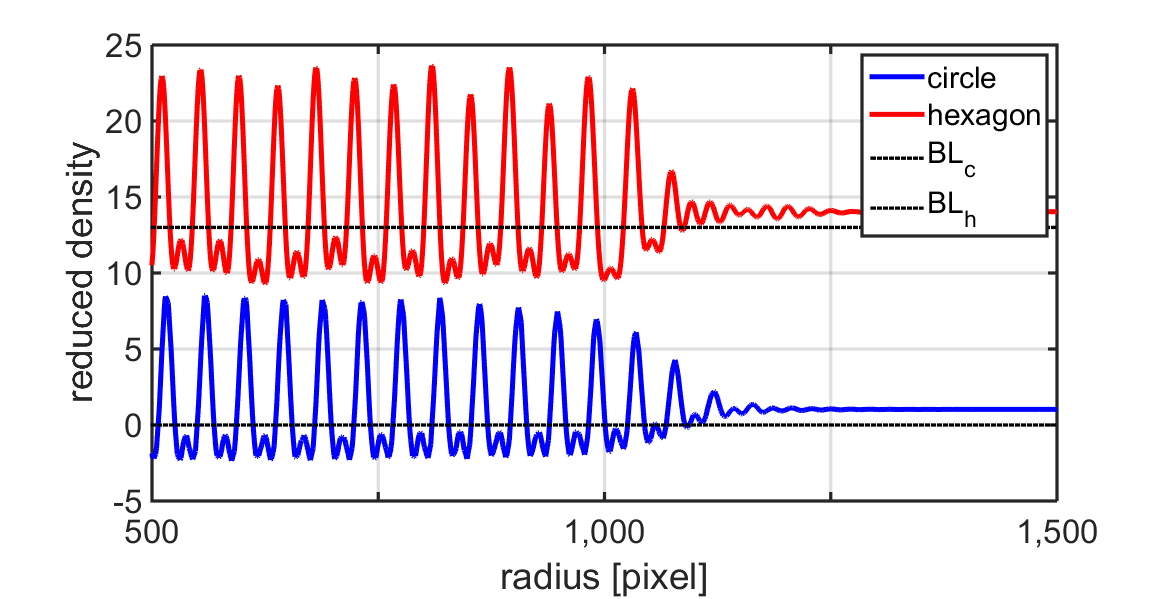}\\ 
(d)\includegraphics[width=7cm]{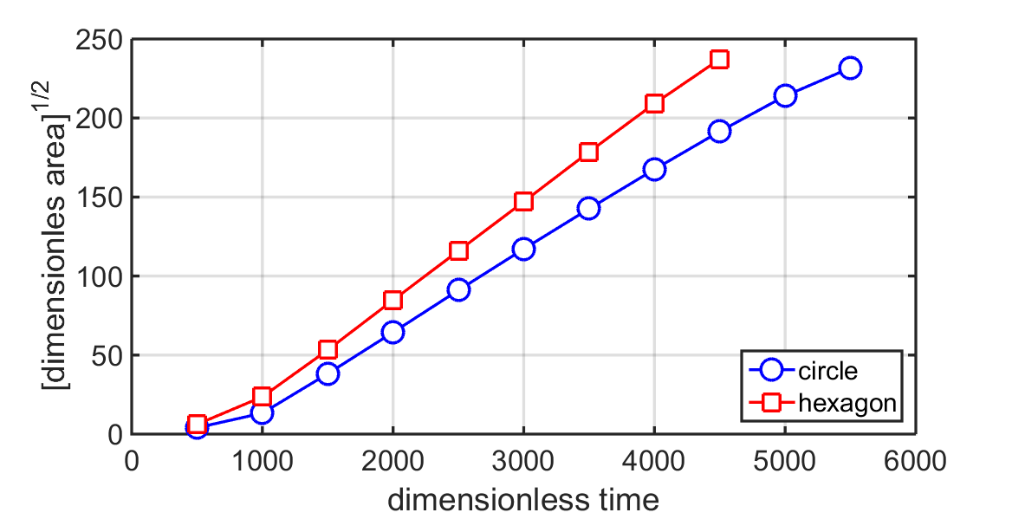}\\ 
\end{tabular}
\caption{(color online) The solid-liquid interface (a) near the liquidus, and (b) close to the liquid stability limit. (c) Interfacial density profile (to enhance the visibility of the density waves, a different coloring is used here.); (d) average radius of the crystal vs. time. In (c) and (d), the upper curves correspond to the hexagonal crystal grown at the stability limit, whereas the lower ones refer to the circular crystal developed near the liquidus. The color scale of panels (a) and (b) was chosen so that it enhances the visibility of the density waves at the solid-liquid interface.   
}
\label{fig:profs}
\end{figure}

The section of the phase diagram, where HPFC simulations were performed is shown in Fig. \ref{fig:PHD}. First, we studied growth initiated by a small (atom size) potential well. Close to the liquidus circular crystals grow, whereas approaching the stability limit, the growth form becomes hexagonal first with rounded corners, evolving into hexagons with pointed corners and concave edges. This behavior is attributed to a change of the interface structure: The respective  density distributions are displayed in Fig. \ref{fig:profs}. Panels (a) and (b) show a closeup of the growth fronts, whereas panels (c) and (d) present the respective density profiles, and the time dependencies of the average crystal radius. Near equilibrium, the solid-liquid interface extends to 6-7 interatomic distances, becoming considerably sharper, when approaching the stability limit. The respective change in anisotropy is expected to be responsible for the different crystal shapes. Apart from an initial transient, the crystals display essentially linear growth as predicted in \cite{ref9}.

\begin{figure}[t]
\begin{tabular}{ccc}
(a)\includegraphics[width=3.cm]{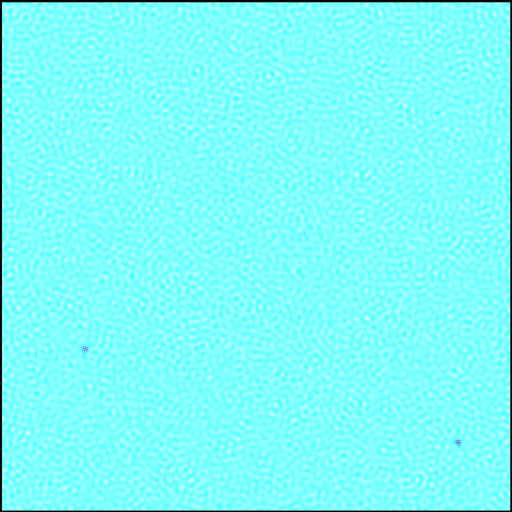} 
(b)\includegraphics[width=3.cm]{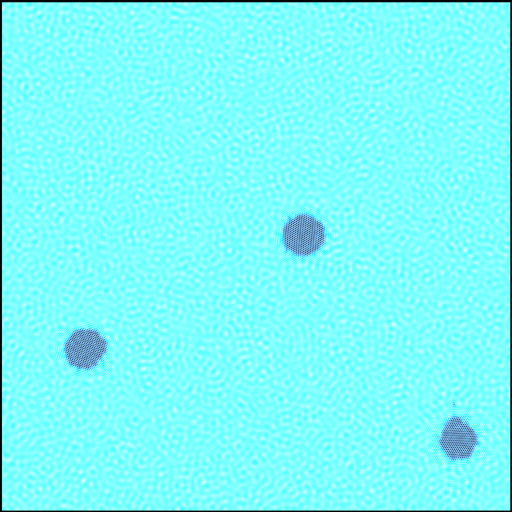}\\ 
(c)\includegraphics[width=3.cm]{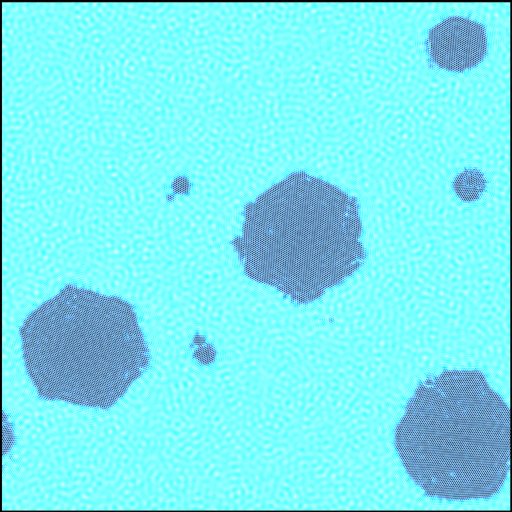} 
(d)\includegraphics[width=3.cm]{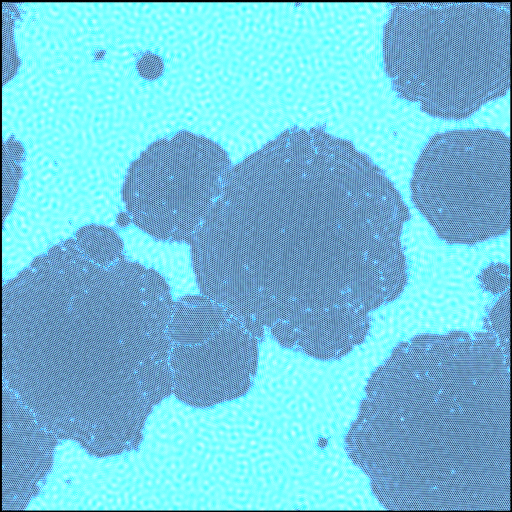}\\ 
(e)\includegraphics[width=3.cm]{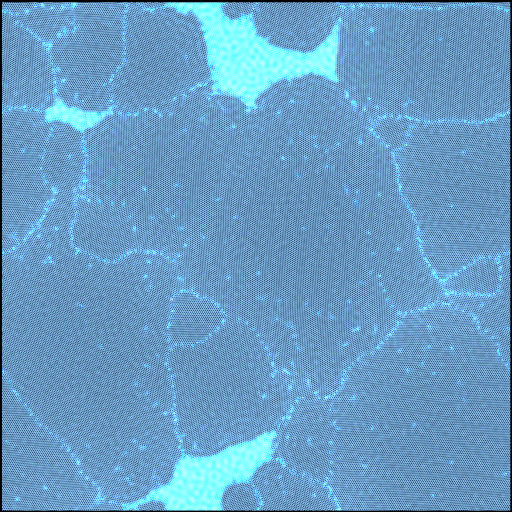} 
(f)\includegraphics[width=3.cm]{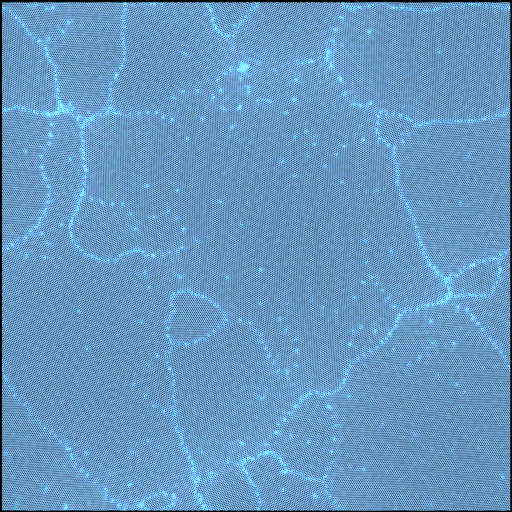}\\ 
(g)\includegraphics[width=7.cm]{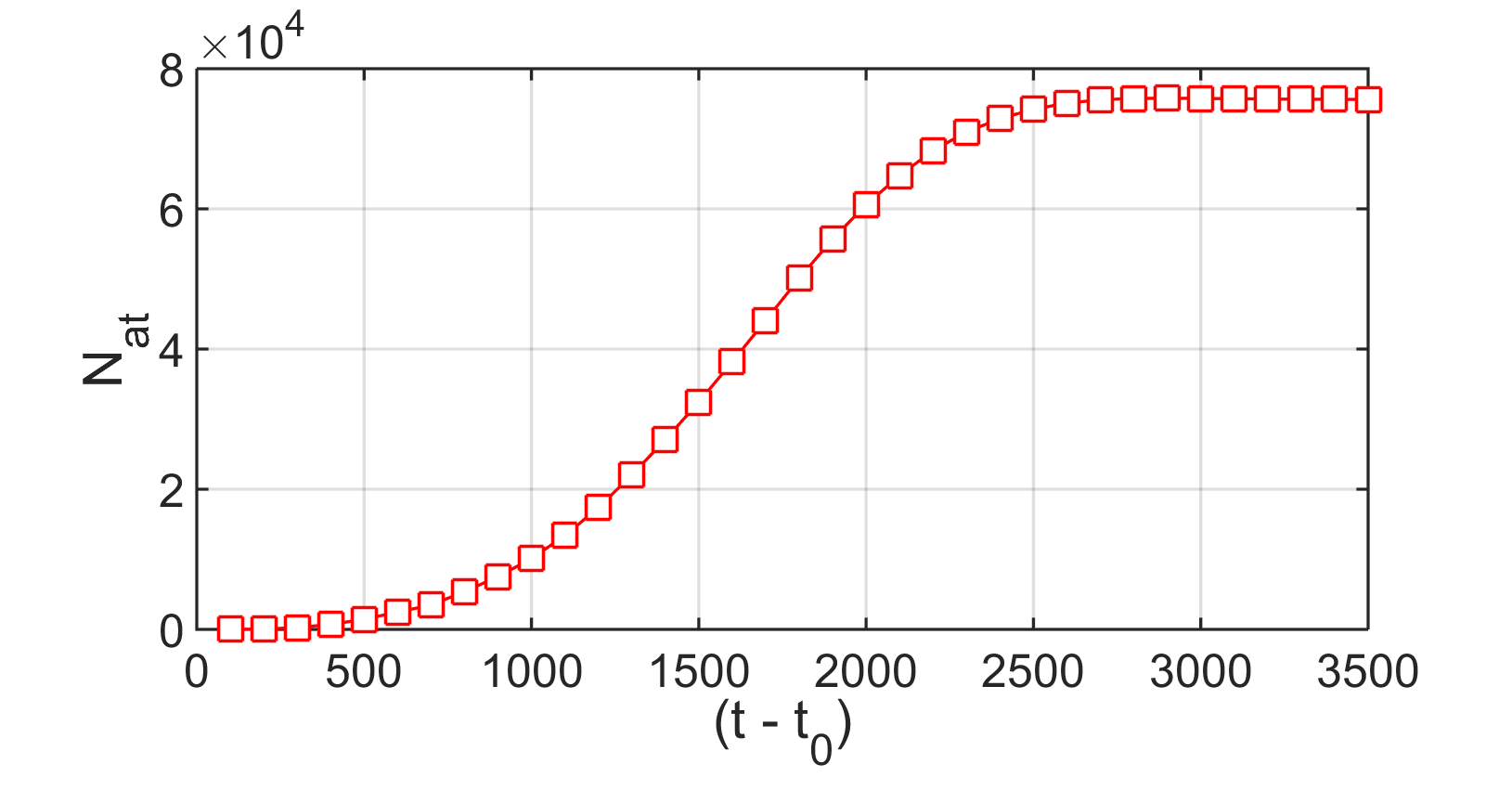}\\ 
(h)\includegraphics[width=7.cm]{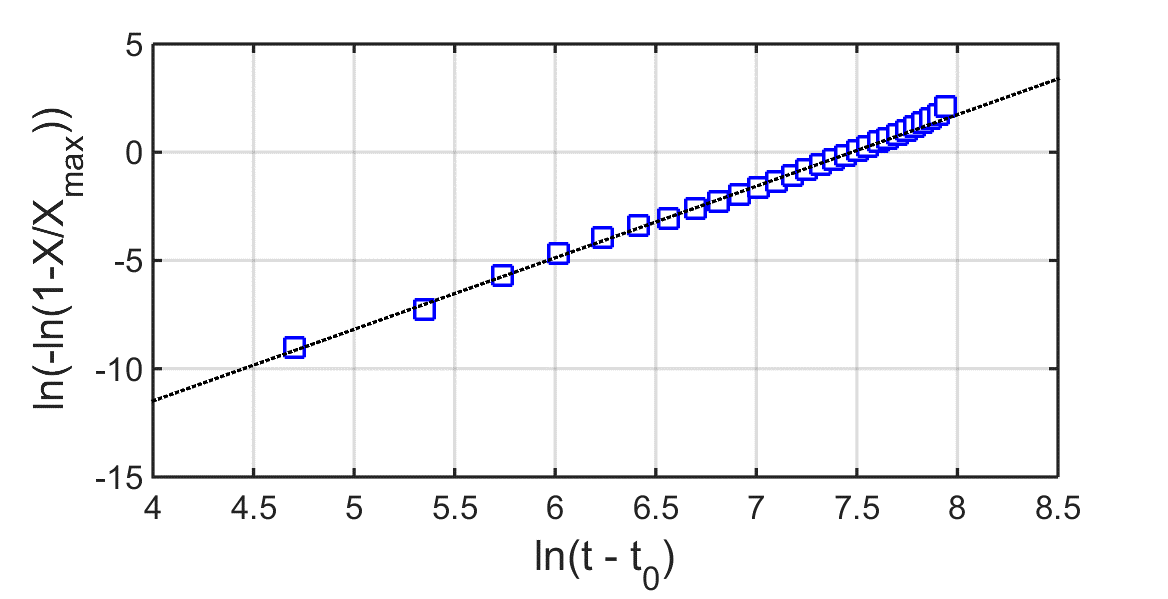}\\ 
\end{tabular}
\caption{(color online) Crystallization kinetics predicted by the HPFC model: (a) -- (f) Snapshots of the density field taken at dimensionless times $t - t_0 = 800, 1100, 1700, 2300, 3000,$ and $3500$, where $t_0 = 3490$ is the incubation time of nucleation. Note the nucleation and growth of crystal grains. (g) Number of atoms in the crystalline phase vs dimensionless time (h) the Avrami plot is nearly linear, yielding $p = 3.32 \pm 0.01$. Here $X/X_{max} = N_{at}/N_{max} = Y$.  
}
\label{fig:JMAK}
\end{figure}

{\it Crystallization kinetics:} The momentum noise in Eq. \ref{eq:NS} gives rise to density fluctuations, which together with molecular scale density waves from the reversible stress tensor, lead to the formation of crystal-like fluctuations (homogeneous nucleation), followed by crystal growth, yielding to polycrystalline freezing. We investigated this in the {\it metastable} liquid domain slightly backward from liquid instability: $\epsilon = 0.1158$ and $\psi_L = - 0.1982$. The results are summarized in Fig. \ref{fig:JMAK}. The Johnson-Mehl-Avrami-Kolmogorov expression, $Y(t) = 1 - exp\{-[(t - t_0)/\tau]^p \}$ \cite{ref2} was fitted to the temperature dependent crystalline fraction evaluated from the number of density peaks. Here $t_0$ is the incubation time, $\tau$ a characteristic time related to the nucleation and growth rates, and $p$ is the Avrami-Kolmogorov exponent indicative to the mechanism of crystallization \cite{ref2}. In our case, from the average slope of the $ln\{-ln(1-Y)\}$ vs. $ln(t - t_0)$ plot, we obtained a $p = 3.31 \pm 0.03$ [Fig. \ref{fig:JMAK}] implying linear growth with slightly increasing nucleation rate. This result differs considerably from the $p$ strongly decreasing with increasing $Y$ reported for diffusive dynamics \cite{ref13}.

\begin{figure}[t]
\begin{tabular}{ccc}
(a)\includegraphics[width=3.cm]{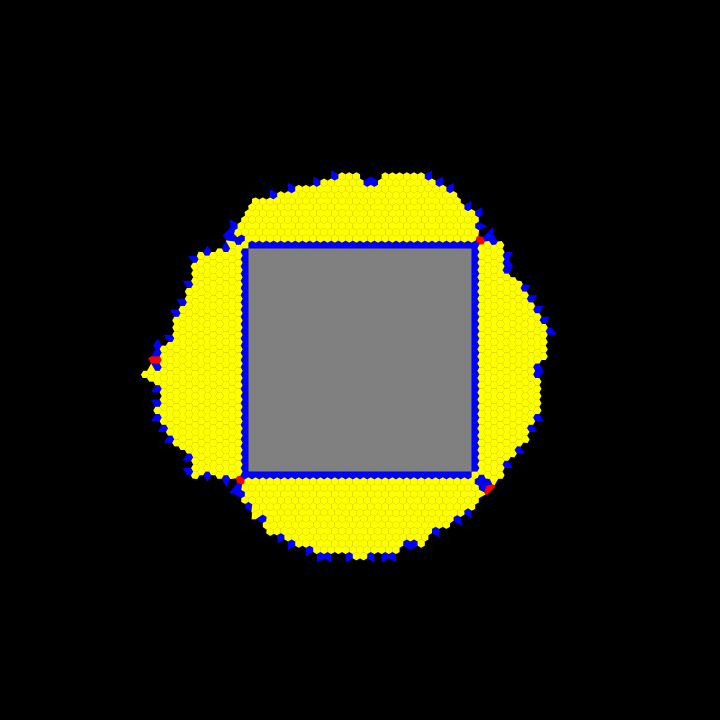} (b)\includegraphics[width=3.cm]{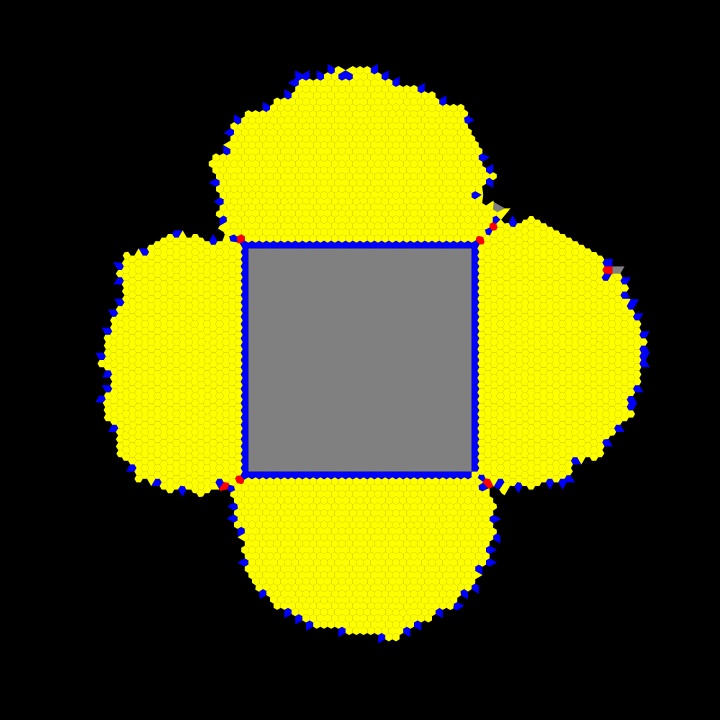}\\
(c)\includegraphics[width=3.cm]{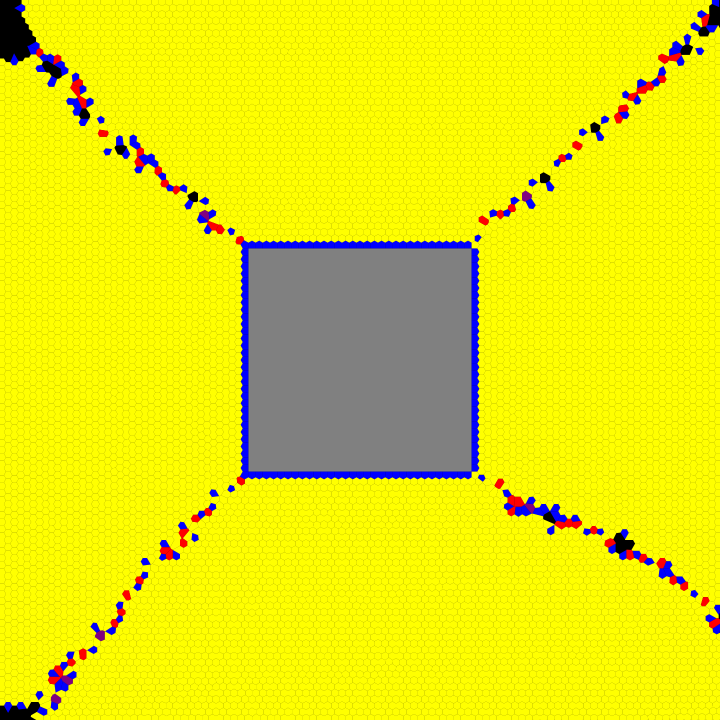} (d)\includegraphics[width=3.cm]{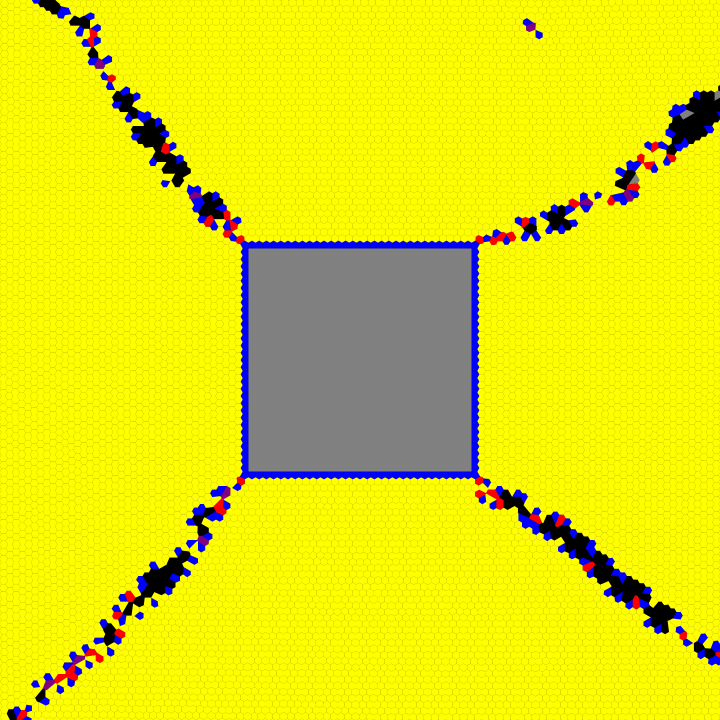}\\ 
\end{tabular}
\caption{(color online) Particle induced freezing: Long-time crystal shapes formed on top of a square substrate as a function of undercooling: (a)-(c) Voronoi maps for the central $800^2$ domain of $2048^2$ simulations made at $\psi = - 0.1982$, and reduced temperatures $\epsilon = 0.0932, 0.0941, 0.970,$ and $0.1017$, respectively are shown.}
\label{fig:free_growth}
\end{figure}

{\it Particle induced 'nucleation':} To model a foreign particle (substrate), an extra term, $V(\mathbf{r})\psi(\mathbf{r}, t)$, was added to the free energy density, where $V(\mathbf{r}) = 0$ outside the substrate, whereas it is a periodic potential inside that determines the crystal lattice of the substrate. To test whether the {\it free growth limited model} of Greer and coworkers \cite{ref14} remains valid in the presence of hydrodynamics, we employ a square shape substrate of made of a square lattice, whose lattice constant coincides with that of the forming triangular crystal. This approximates the ideal wetting Greer and coworkers assumed between the substrate and the crystal. The undercooling has been increased by multiplying $C_0$ by factors $\xi = 0.999, 0.998, 0.995,$ and $0.990$. In Fig. \ref{fig:free_growth}, we present long-time solutions as a function of $\xi C_0$. In the HPFC model (as well as in the PFC model with diffusive dynamics \cite{ref15}), there exists a critical undercooling for a given particle size, beyond which free growth takes place.

\begin{figure}[t]
\begin{tabular}{ccc}
\includegraphics[width=7.cm]{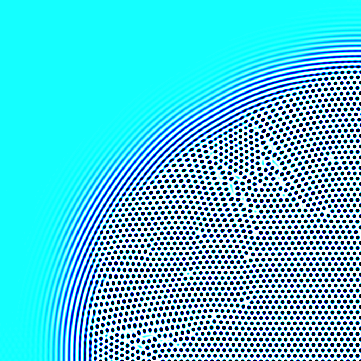}\\
\end{tabular}
\caption{ (color online) Density waves, and the formation of defects and new crystallographic orientations {\it beyond the linear stability limit of the liquid} in the PFC model with diffusive dynamics \cite{ref16}. The amplitude of the noise was set to zero, while the other parameters were chosen as follows: $\psi_0 = -0.45$, $\epsilon = 0.75$. The reduced density corresponding to linear stability at this $\epsilon$ is $\psi_c = -0.5$. The upper left quarter of a $2048^2$ simulation is shown. The color scale was chosen so that it enhances the visibility of the density waves at the solid-liquid interface.}
\label{fig:dpfc}
\end{figure}

{\it Growth Front Nucleation (GFN):} Formation of new grains at the solidification front has been identified as the mechanism by which complex polycrystalline growth forms appear \cite{ref3,ref4,ref5}. This phenomenon has been successfully modeled by phase-field methods employing orientation fields to monitor the local crystallographic orientation. In these models, new grains form either by quenching orientational defects (bundles {\color{blue} of} dislocations) into the crystal (at large undercoolings), or via branching in directions of low grain-boundary energies (at small undercoolings). The orientation field approach, became fairly successful in capturing complex growth structures \cite{ref5}. Yet it is desirable to clarify the microscopic background of GFN. It is suspected that in substances of different molecular structures/interactions different GFN mechanisms may occur. Herein, we employ the HPFC model to study the formation of new grains at the solid-liquid interface of 2D hexagonal crystals {\it at high undercoolings}.

\begin{figure}[t]
\begin{tabular}{ccc}
(a)\includegraphics[width=2.5cm]{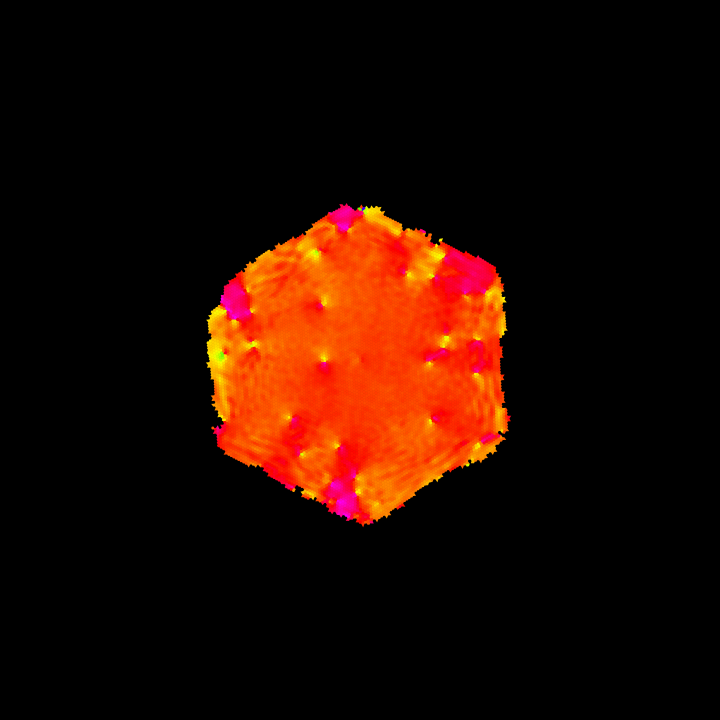} (b)\includegraphics[width=2.5cm]{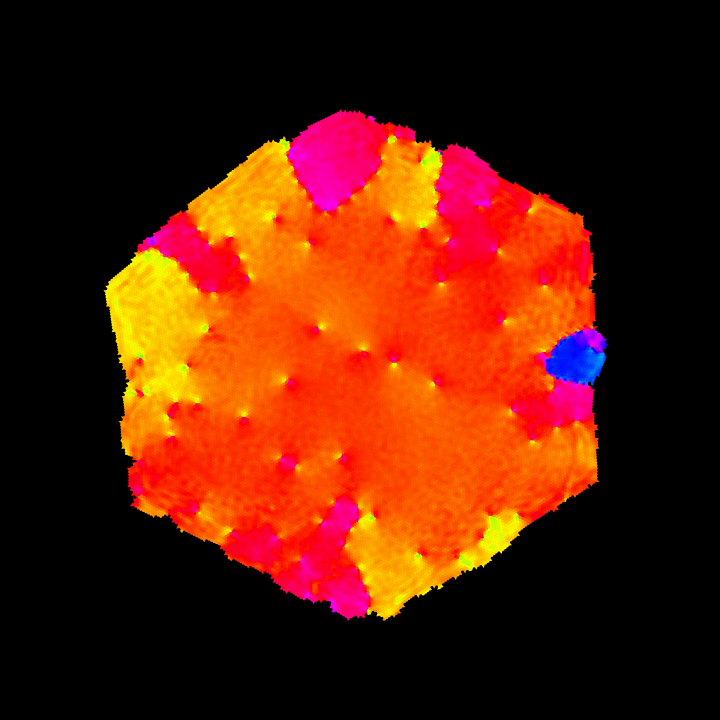} (c)\includegraphics[width=2.5cm]{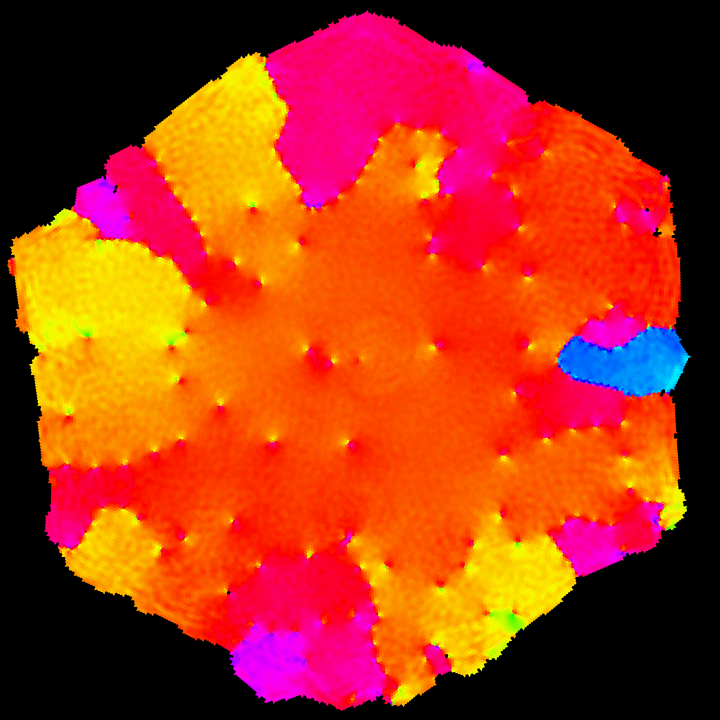}\\
(d)\includegraphics[width=2.5cm]{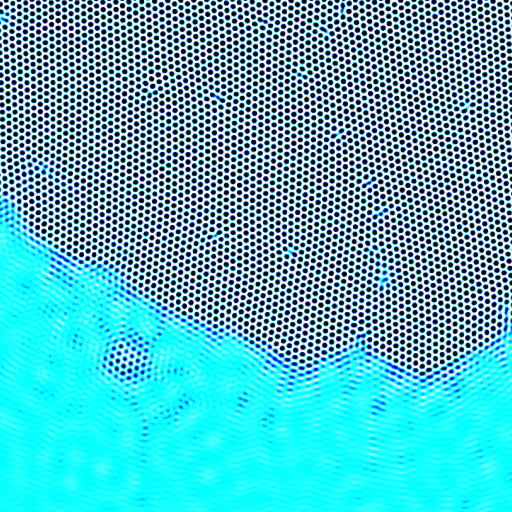} (e)\includegraphics[width=2.5cm]{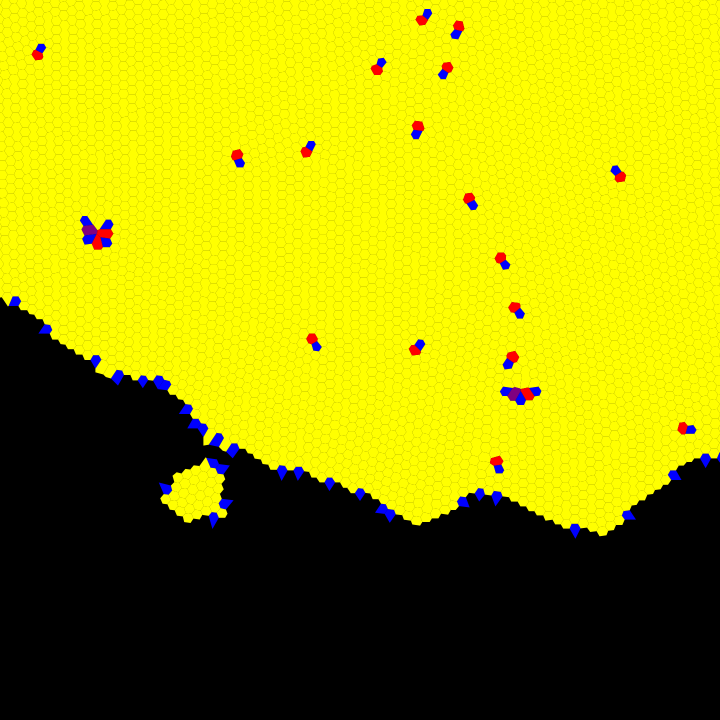} (f)\includegraphics[width=2.5cm]{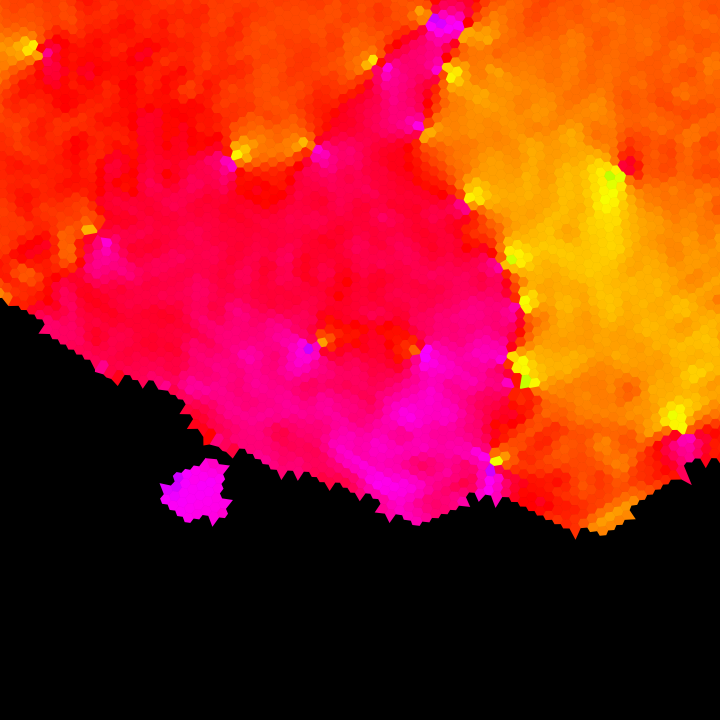}\\
(g)\includegraphics[width=2.5cm]{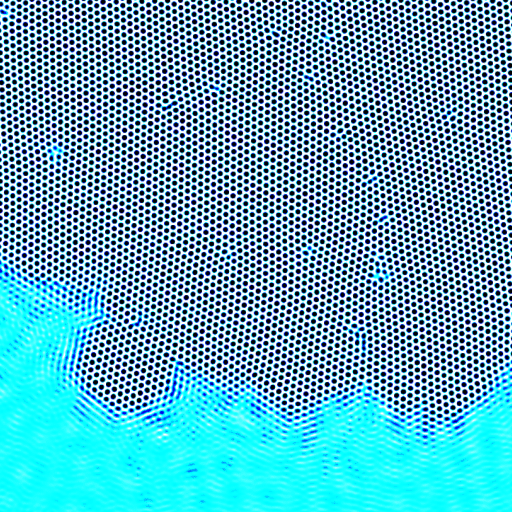} (h)\includegraphics[width=2.5cm]{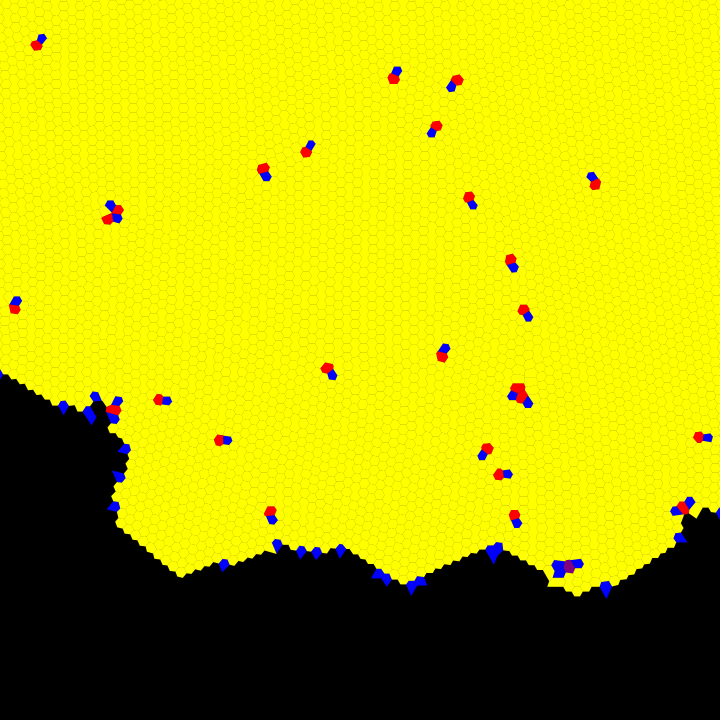} (i)\includegraphics[width=2.5cm]{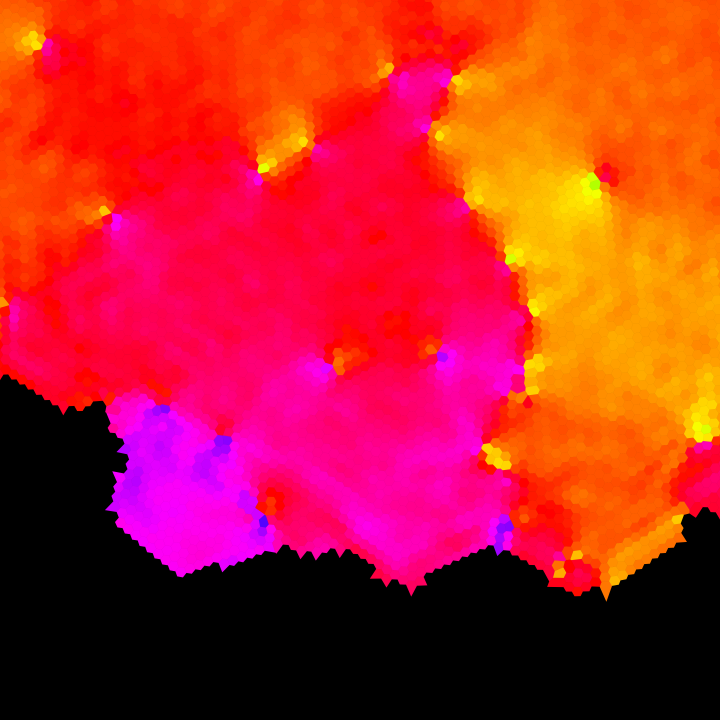}\\
(j)\includegraphics[width=2.5cm]{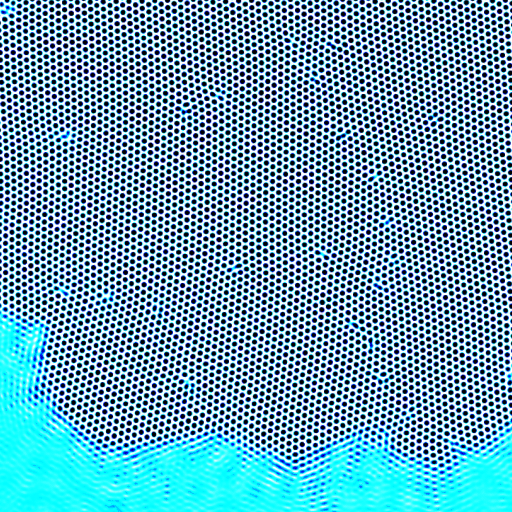} (k)\includegraphics[width=2.5cm]{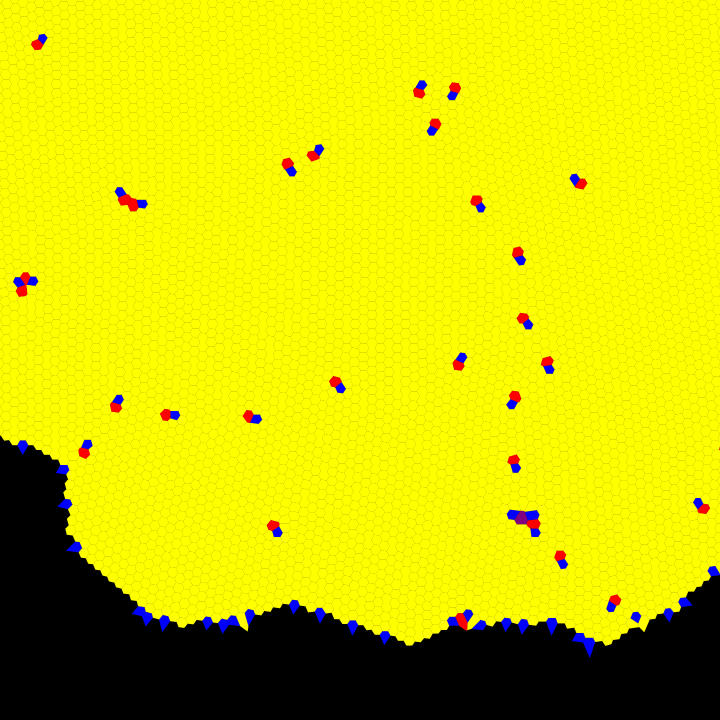} (l)\includegraphics[width=2.5cm]{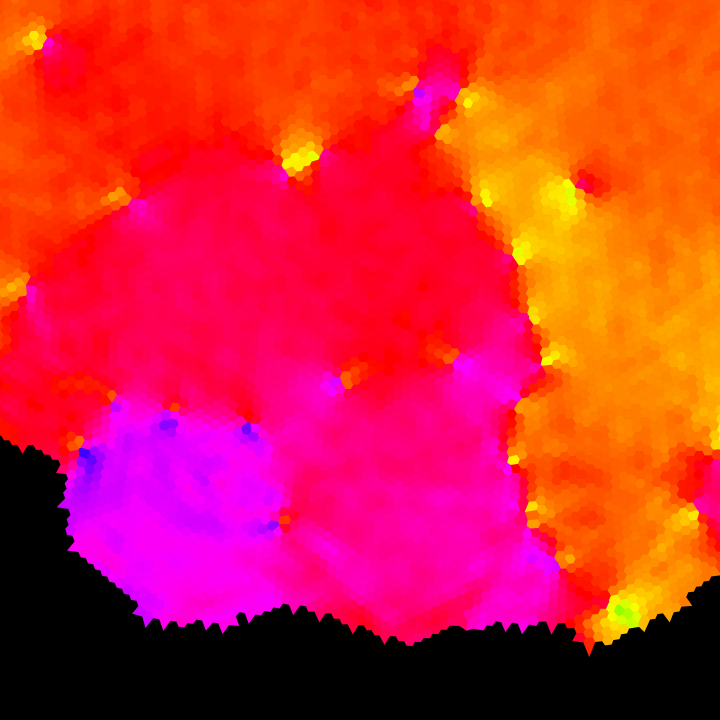}\\

\end{tabular}
\caption{(color online) Polycrystalline growth in the metastable liquid of scaled density $\psi_0 = - 0.1982$, close to the liquid stability limit ($\epsilon = 0.1158 < \epsilon_c = 0.1178$). Two mechanisms of GFN are observed: (i) Formation of dislocation chains (chains of blue-red pairs) due to the interaction of local stresses with density fluctuations, and (ii) formation of nuclei close to the growth front due to density waves emanating from the  rough solid-liquid interface: (a)-(c) Orientation map vs. time ($t = 2100, 2900,$ and $3900$; $2048^2$ grid); (d)-(l) density, Voronoi, and orientation maps showing the two GFN modes ($600^2$ segment). The color scale of panels (d), (g), and (h) was chosen so that it enhances the visibility of the density waves at the solid-liquid interface. (For computer animations see http://phasefield.hu/pages/2017pre/)}
\label{fig:GFN1}
\end{figure}

We made our first attempts to model GFN years ago \cite{ref16} using the original PFC model. At supersaturations beyond the liquid stability limit $[\psi > \psi_c = - (\epsilon/3)^{1/2}]$, we observed that {\it without noise} crystal seeds evolved into ordered polycrystalline structures. The growing crystal was surrounded by concentric density waves, which initiated crystallization accordingly: in six directions these waves helped the growth of the original crystal, whereas in other directions a large number of defects formed and new orientations appeared that fitted better to the local direction of the density waves {\color{blue}(Fig. \ref{fig:dpfc})}. Later works performed {\it without noise} in the unstable liquid regime indicate the transition of a flat single crystal front into polycrystalline growth and eventually into glassy freezing with increasing driving force \cite{ref17}. These findings could be regarded as signs of elementary processes of GFN. However, when a noise term obeying the fluctuation-dissipation theorem is added to the equations of motion in the unstable regime, these solutions are suppressed by explosive nucleation. Inside the {\it metastable} regime, we were unable to observe GFN, probably owing to the lack of an extended layer of molecular scale density waves ahead of the front.

\begin{figure}[t]
\begin{tabular}{ccc}
\includegraphics[width=2.5cm]{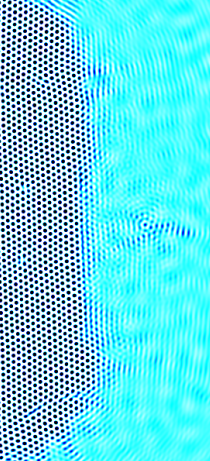}
\includegraphics[width=2.5cm]{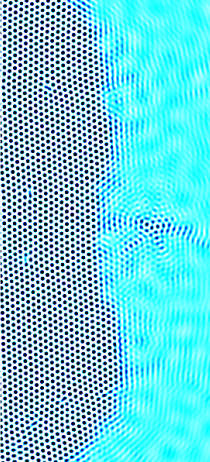}
\includegraphics[width=2.5cm]{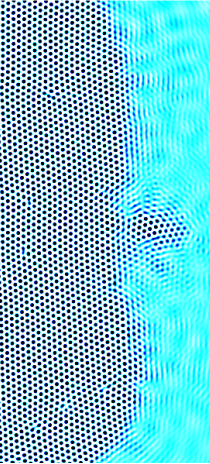}
\end{tabular}
\caption{ (color online) Crystal nucleation initiated by the interference of density waves ahead of the solidification front on the right hand side of the crystal shown in Fig. \ref{fig:GFN1}: Snapshots of the particle density field were taken at dimensionless times $t = 1900, 2000$, and $2100${\color{blue}. omit ''that show''} The color scale was chosen so that it enhances the visibility of the density waves at the solid-liquid interface. (For computer animations see http://phasefield.hu/pages/2017pre/)}
\label{fig:GFN2}
\end{figure}

In the HPFC model, we were able to observe polycrystalline growth forms in the {\it metastable liquid} (see Fig. \ref{fig:GFN1}). The kinetic equations were solved on a $2048^2$ rectangular grid. A shallow molecular size potential well was used to initiate freezing. It induced concentric density waves (akin to the 'onion structures' predicted in \cite{ref18}, then a small hexagonal single crystal formed, but as it grew new orientations appeared gradually via two mechanisms of GFN: (i) Dislocations entered in the growing crystallites at cusps centers. These appear due to the interaction of the stress field of the growing nanoscale crystallite with density fluctuations. (ii) Small crystallites nucleate in the close vicinity of the solid-liquid interface, which apparently originate from the interference of the density waves ahead of the the rough solid-liquid interface (see Fig.  \ref{fig:GFN2}). The two mechanisms are clearly visible in the snapshots of the density, orientation and Voronoi maps (Figs. \ref{fig:GFN1} and \ref{fig:GFN2}).

We found that mechanism (i) occurs in a relatively broad range of undercoolings or densities, however, the formation rate of dislocations decreases with decreasing driving force, a finding that might be associated with a decreasing growth velocity (see the discussion below). In contrast, mechanism (ii) can only be observed in the close vicinity of the linear stability limit of the liquid. For example, at the reduced temperature $\epsilon = 0.0923$, mechanism (ii) appears in the scaled density range $-0.1778 < \psi_0 < \psi_c = -0.1754$; whereas at fixed scaled density of $\psi_0 = - 0.1982$, nucleation ahead of the front occurs in the reduced temperature range $ 0.1158 < \epsilon < \epsilon_c = 0.1178$. This observation correlates with the finding that in the HPFC model, the thickness of the liquid layer, in which liquid ordering in the form of density waves takes place, increases towards the stability limit.   

\begin{figure}[t]
\includegraphics[width=7.cm]{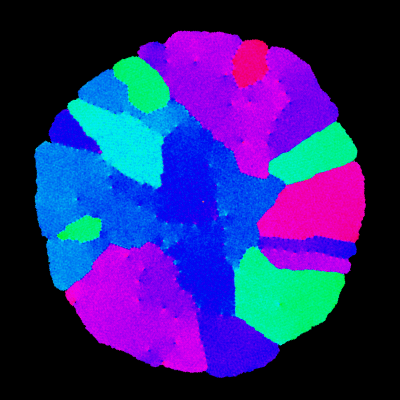}
\caption{(color online) Snapshot of the orientation field in a larger scale $(8192^2)$ simulation of GFN for $\epsilon = 0.1111$ and $\psi = - 0.1982$. Note the spherulite-like morphology. 
}
\label{fig:spherulite}
\end{figure}

Although our simulations are on the nanoscale, we observe spherulite-like structures (Fig. \ref{fig:spherulite}) similar to those seen on larger scales. We note that polycrystalline growth structures can form spontaneously on this scale as shown by molecular dynamics simulations \cite{ref19} and experiments on organic and metallic nanospherulites \cite{ref20}. Our work might be directly relevant to them and to carbon nanostructures.

An intriguing question is, why the HPFC model recovers GFN in the metastable liquid domain, whereas the diffusive PFC model does not. This might be related to either the differences in the interface structure or defect dynamics, or in both. For example, in the case of diffusive PFC, a fast diffusionless growth mode characterized by a broad interface occurs at large driving forces (still in the metastable liquid) \cite{ref21}, in which healing of the defects can be relatively easy, avoiding the formation of dislocations at the perimeter of the growing crystal, preventing thus GFN. In the HPFC model, the relatively sharp inter face may be unable to do this. We discuss these issues below. 

The original PFC model incorporates elastic interactions as well as crystal plasticity combined with diffusive dynamics, however without hydrodynamic modes of the evolution of the density field. Defect dynamics, including vacancy motion, dislocation glide, climb, and annihilation, and grain boundary melting were studied extensively on the diffusive time scale \cite{ref7,ref8,ref22}. A few studies went beyond this approximation by incorporating faster processes enabled by the inclusion of a second order time derivative into the equation of motion (see the MPFC model in Ref. \cite{ref23}), or via linearizing the Navier-Stokes equations \cite{ref24}. In both of these quasi-hydrodynamic models the atomic positions are relaxed rapidly at early times in a manner consistent with elasticity theory, while the late time defect motion, including vacancy diffusion, grain boundary kinetics, and dislocation climb, is governed by diffusive dynamics. This suggests that defect dynamics alone may not be responsible for the observed differences in GFN. 

A work by Majaniemi {\it et al.} \cite{ref24} explored differences between dynamics of mass distribution in the MPFC model (they termed it Type-2 model) and a more sophisticated linearized hydrodynamic model (termed Type-3 model) in the non-equilibrium case of a crack relaxation under stress. They found that the dynamic mass distribution the two models predict can be quite different, which they attributed to the transport differences of the quasi-phonons appearing in the MPFC model and the acoustic phonons of the Type-3 model. Remarkably, in the Type-3 linearized hydrodynamics model, which is close to the present full hydrodynamic model, the system relaxed to a multigrain structure, as opposed to the single crystal solution from the MPFC model. The behavior predicted by the Type-3 model appears to be similar to what we observed in our full hydrodynamic HPFC simulations, in which, after engulfing low density fluctuations into the solid, dislocations and new grains were formed at the interface. This phenomenon is more frequent in the cusps of the interface, and the dislocations may be of misfit origin.

\begin{figure}[t]
\begin{tabular}{ccc}
(a)\includegraphics[width=3.5cm]{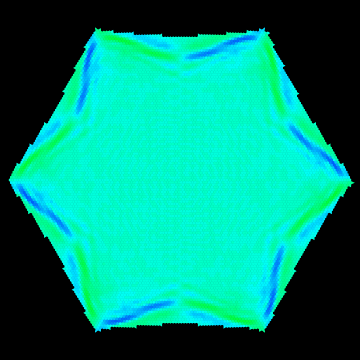}
(b)\includegraphics[width=4cm]{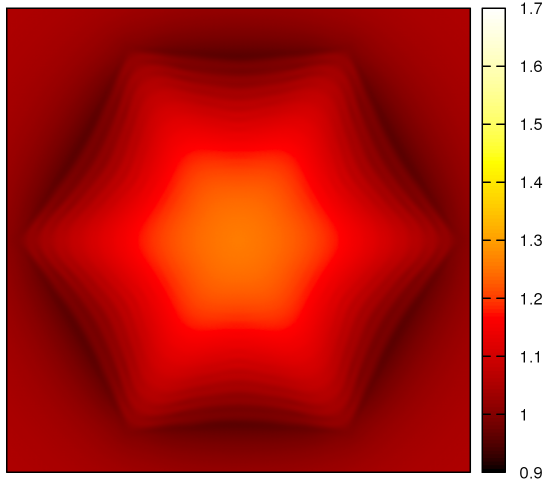} 
\end{tabular}
\caption{ (color online)  Single crystal growing {\it without fluctuations} at $\epsilon = 0.1158$ and $\psi = - 0.1982$. (a) Snapshot of the central $1024^2$ section of the orientation field in a $2048^2$ simulation, taken at dimensionless time $t = 4000$. (b) Snapshot of coarse grained (FIR filtered) density for the same area.   
}
\label{fig:orinonois}
\end{figure}

\begin{figure}[t]
\begin{tabular}{ccc}
(a)\includegraphics[width=4cm]{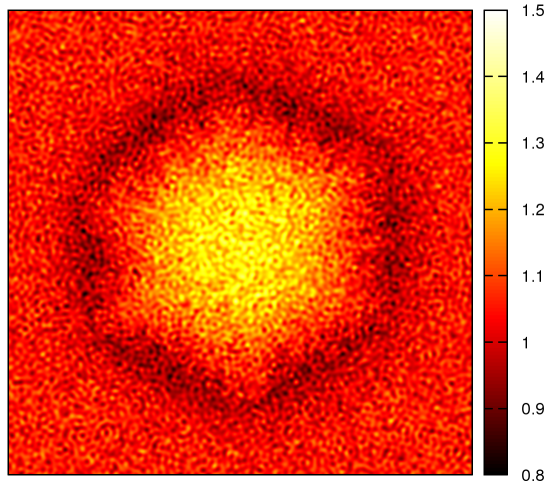}
(b)\includegraphics[width=3.5cm]{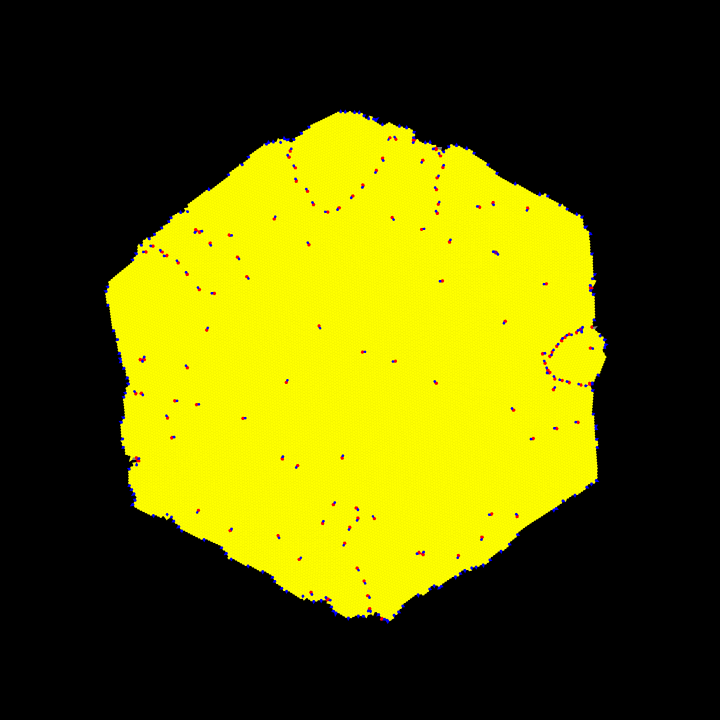} 
\end{tabular}
\caption{ (color online) Crystallite growing {\it in the presence of momentum fluctuations} at $\epsilon = 0.1158$ and $\psi = - 0.1982$, in a $2048^2$ simulation. (a) Snapshot of coarse grained (FIR filtered) density, taken at dimensionless time $t = 2900$. (b) Snapshot of the Voronoi map at the same time.    
}
\label{fig:noidens}
\end{figure}

Apparently, we have stress at the growth front already in the early single crystal state, as reflected by the non-uniform orientation field that indicates slight misorientations (relative rotation of the Voronoi cells) between the two sides of the corners of the crystallite and at the central region of its sides. To see this clearly, we performed simulations without momentum fluctuations for the highly nonequilibrium liquid, starting crystallization with a small cluster that evolves due to a weak potential of 2D hexagonal structure. The potential contains a central well and 6 neighboring wells of equal depth given by the single mode solution for the particle density for this structure. 

The orientation field of a crystallite, grown from such a cluster, is shown in Fig. \ref{fig:orinonois}(a). The deformation and the related stress that causes the inhomogeneity of the orientation field appears to originate from a radial density change shown in Fig. \ref{fig:orinonois}(b). The latter panel displays the coarse grained particle density obtained by FIR (Finite Impulse Response) filtering \cite{ref25} from the particle density. The spatial modulation of coarse grained density at the interface might indicate a weakly oscillating growth velocity. Despite the presence of the hydrodynamic mode of density relaxation, {\it density depletion} is observed ahead of the solidification front, which is probably responsible for the curving of the sides of the crystal, yielding orientation and stress fields antisymmetric to the lines across the tips and centers of the opposite sides. An essentially similar, though more noisy coarse grained distribution is obtained in the presence of momentum fluctuations [see Fig. \ref{fig:noidens}(a)].

We observe the formation of dislocations at the interface both with [Fig. \ref{fig:noidens}(b)] and without momentum fluctuations, however at a much later stage in the latter case: In the simulations without noise, we do not observe dislocations below an equivalent radius ($R_c = 2\sqrt{2 \pi N/3} \approx 450$, where $N$ is the number of particles in the crystallite, see Figs. \ref{fig:orinonois}). Beyond this size the stress at the interface is large enough to initiate the formation of cusps, and the nucleation of dislocations in them. A possible mechanism for this can be the Asaro-Tiller-Grinfeld instability \cite{ref26,ref27}. We find that these misfit dislocations appear at a much smaller size in the presence of momentum noise ($R_c \approx 150$). These findings indicate that the fluctuations play a key role in the formation of defects. This conclusion is further supported by a recent PFC study, which has shown that the formation of misfit dislocations is helped by an increasing strength of the density fluctuations \cite{ref27}. We note furthermore, that the noise influences the appearance of dislocations, which leads to symmetry breaking \cite{ref27}. Apparently, in the presence of density fluctuations the dislocations appear fairly randomly, although normally in cusps forming at the interface.

To test the role of fluctuations further, we have determined the wavenumber spectrum of the density fluctuations in the HPFC model emerging from the momentum fluctuations, and included the same noise in the equation of motion of the diffusive PFC model under the same $\epsilon$ and $\psi_0$ values (under these conditions the diffusionless 'fast growth mode', takes place in the diffusive PFC model \cite{ref21}). Despite these, no GFN was observed in the diffusive PFC model. This observation suggests that under equivalent conditions the diffusive PFC model is less susceptible to defect formation than the HPFC model. 

It appears that a combination of faceted growth, a weak depletion at the interface, and the presence of density fluctuations is needed for initiating GFN. Work is underway to investigate further the microscopic aspects of these phenomena.

Summarizing, we applied the HPFC model to solidification problems in 2D. We demonstrated that
 
	(i) radial growth happens at a steady state rate,

	(ii) crystallization takes place via homogeneous nucleation and steady state growth, and the respective Avrami-Kolmogorov exponent is $p = 3.31 \pm 0.03$,   

	(iii) the free growth limited mode of particle induced crystallization \cite{ref14} remains valid down to the nanoscale for simple liquids, and 

	(iv) the model predicts two modes for growth front nucleation.     

While (i) to (iii) are important demonstrations of the capabilities of the model, the most significant result is (iv): The HPFC appears to be the first atomic scale model that yields growth front nucleation in the metastable liquid regime. The identification of the basic mechanisms of growth front nucleation in simple liquids might help to control microstructure evolution in such systems.

\begin{acknowledgments} This work has been supported by NKFIH, Hungary under contract No OTKA-K-115959, by the ESA MAP/PECS projects “MAGNEPHAS III” (Contract No 40000110756/11/NL/KML) and “GRADECET” (Contract No 40000110759/11/NL/KML). G. Tegze is a grantee of the J\'anos Bolyai Scholarship of the MTA, Hungary.
\end{acknowledgments}


\begin{thebibliography}{99}

\bibitem{ref1} K. F. Kelton and A. L. Greer, \textit{Nucleation in condensed matter} (Elsevier, Amsterdam, 2010).

\bibitem{ref2} (a) J. W. Christian, {\it Transformations in Metals and Alloys} (Pergamon, Oxford, 1981); (b) W. Kurz and D. J. Fisher, {\it Fundamentals of Solidification} (Trans. Tech. Publ., Aedermannsdorff, 1985).

\bibitem{ref3} V. Ferreiro, J. F. Douglas, J. A. Warren, and A. Karim, Phys. Rev. E {\bf 65}, 042802 (2002); V. Ferreiro, J. F. Douglas, J. A. Warren, and A. Karim, Phys. Rev. E {\bf 65}, 051606 (2002). 

\bibitem{ref4} J. H. Magill, J. Mater. Sci. {\bf 36}, 3143 (2001). 

\bibitem{ref5} L. Gr\'an\'asy, T. Pusztai, J. A. Warren, J. F. Douglas, T. B\"orzs\"onyi, and V. Ferreiro: Nature Mater. {\bf 2}, 92 (2003); L. Gr\'an\'asy, T. Pusztai, T. B\"orzs\"onyi, J. A. Warren, J. F. Douglas, Nature Mater. {\bf 3}, 645 (2004); L. Gr\'an\'asy, T. Pusztai, G. Tegze, J. A. Warren, and J. F. Douglas, Phys. Rev. E {\bf 72}, 011605 (2005); L. Gr\'an\'asy, L. R\'atkai, A. Sz\'all\'as, B. Korbuly, G. I. T\'oth, L. K\"ornyei, and T. Pusztai, Metall. Mater. Trans. A {\bf 45}, 1694 (2014). 

\bibitem{ref6} P. R. ten Wolde, M. J. Ruiz-Montero, and D. Frenkel Phys. Rev. Lett. {\bf 75}, 2714 (1995); J. F. Lutsko and G. Nicolis, Phys. Rev. Lett. {\bf 96}, 046102 (2006); T. Schilling, H. J. Sch\"ope, M. Oettel, G. Opletal, and I. Snook, Phys. Rev. Lett. {\bf 105}, 027501 (2010); T. Kawasaki and H. Tanaka, Proc. Natl. Acad. Sci. {\bf 107}, 14036 (2010); H. Tanaka, Eur. Phys. J. E {\bf 35}, 1 (2012); G. I. T\'oth, T. Pusztai, G. Tegze, G. T\'oth, and L. Gr\'an\'asy, Phys. Rev. Lett. {\bf 107}, 175702 (2011).  

\bibitem{ref7} K. R. Elder, M. Katakowski, M. Haataja, and M. Grant, Phys. Rev. Lett. {\bf 88}, 245701 (2002); K. R. Elder, N. Provatas, J. Berry, P. Stefanovic, and M. Grant, Phys. Rev. B {\bf 75}, 064107 (2007); K.-A. Wu, A. Adland, and A. Karma, Phys. Rev. E {\bf 81}, 061601 (2010); M. Greenwood, N. Provatas, and J. Rotter, Phys. Rev. Lett. {\bf 105}, 045702 (2010). 

\bibitem{ref8} For a recent review on PFC models see: H. Emmerich, H. L\"owen, R. Wittkowski, T. Gruhn, G. I. T\'oth, G. Tegze,  and L. Gr\'an\'asy, Adv. Phys. {\bf 61}, 665 (2012).

\bibitem{ref9} G. I. T\'oth, L. Gr\'an\'asy, and G. Tegze, J. Phys.: Condens. Matter {\bf 26}, 055001 (2014).

\bibitem{ref10} B. Z. Shang, N. K. Voulgarakis, and J.-W. Chu, J. Chem. Phys. {\bf 135}, 044111 (2011).

\bibitem{ref11} A. Baskaran, A. Baskaran, and J. Lowengrub, J. Chem. Phys. {\bf 141}, 174506 (2014); V. Heinonen, C. V. Achim, J. M. Kosterlitz, S.-C. Ying, J. Lowengrub, and T. Ala-Nissila, Phys. Rev. Lett. {\bf 116}, 024303 (2016); A. Baskaran, Z. Guan, and J. Lowengrub, Comput. Meth. Appl. Mech. Eng. {\bf 299},  22 (2016). 

\bibitem{ref12} H. Okumura, Mol. Phys. {\bf 104}, 3751 (2006); G. De Fabritiis, M. Serrano, R. Delgado-Buscalioni, and P. V. Coveney, Phys. Rev. E {\bf 75}, 026307 (2007); R. Delgado-Buscalioni, E. Chac\'on, and P. Tarazona, J. Phys.: Condens. Matter {\bf 20}, 494229 (2008); N. K. Voulgarakis, S. Satish, and J.-W. Chu, J. Chem. Phys. {\bf 131}, 234115 (2009); A. Markesteijn, S. Karabasov, A. Scukins, D. Nerukh, V. Glotov, and V. Golovizin, Phil. Trans. R. Soc. A {\bf 372}, 20130379 (2014).

\bibitem{ref13} E. Pineda, T. Pradell. D. Crespo, N. Clavaguera, and M. T. Clavaguera-Mora, J. Non-Cryst. Solids {\bf 287}, 92 (2001); L. Gr\'an\'asy, T. B\"orzs\"onyi, and T. Pusztai, Phys. Rev. Lett. {\bf 88}, 206105 (2002); T. Pusztai, G. Tegze, G. I. T\'oth, L. K\"ornyei, G. Bansel, Z. Fan, and L. Gr\'an\'asy, J. Phys.: Condens. Matter {\bf 20}, 404205 (2008).

\bibitem{ref14} A. L. Greer, A. M. Brunn, A. Tronche, P. V. Evans, and D. J. Bristow, 
Acta Mater. \textbf{48}, 2823 (2000); T. E. Quested and A. L. Greer, Acta Mater. \textbf{53}, 2683 (2005).

\bibitem{ref15} G. I. T\'oth, G. Tegze, T. Pusztai, and L. Gr\'an\'asy, Phys. Rev. Lett. {\bf 108}, 025502 (2012). 

\bibitem{ref16} L. Gr\'an\'asy, G. Tegze, G. I. T\'oth, F. Podmaniczky, and T. Pusztai, Atomistic phase-field approach to crystal nucleation and growth in two and three dimensions. {\it GRC on Thin Films and Crystal Growth Mechanisms}, 12-17 July 2009, New London, NH.

\bibitem{ref17} A. J. Archer, M. J. Robbins, U. Thiele, and E. Knobloch, Phys. Rev. E {\bf 86}, 031603 (2012); A. J. Archer, M. C. Walters, U. Thiele, and E. Knobloch, Phys. Rev. E {\bf 60}, 042404 (2014).

\bibitem{ref18} K. Barros and W. Klein, J. Chem. Phys. {\bf 139}, 174505 (2013).

\bibitem{ref19} B. O'Malley and I. Snook, Phys. Rev. Lett. {\bf 90}, 085702 (2003); Ph. Lambin, A. Loiseau, M. Monthioux, and J. Thibault, in {\it  Understanding Carbon Nanotubes}, eds. A. Loiseau, P. Launois, P. Petit, S. Roche, and J.-P. Salvetat (Springer, Berlin, Heidelberg, 2006), Chapt. 3, p. 161; T. Koishi, K. Yasuoka, and T. Ebisuzaki, J. Chem. Phys. {\bf 119}, 11298 (2003).

\bibitem{ref20} T. Yano, Y. Yorikado, Y. Akeno, F. Hori, Y. Yokoyama, A. Iwase, A. Inoue, and T. J. Konno, Mater. Trans. {\bf 46}, 2886 (2005); P. J. Phillips, IEEE Trans. Electr. Insul. {\bf EI 13}, 69 (1978); B. Fitton and C. H. Griffiths, J. Appl. Phys. {\bf 39}, 3663 (1968).

\bibitem{ref21} G. Tegze, G. I. T\'oth, and L. Gr\'an\'asy, Phys. Rev. Lett. {\bf 106}, 195502 (2011); G. Tegze, L. Gr\'an\'asy, G. I. T\'oth, J. F. Douglas, and T. Pusztai, Soft Matter {\bf 7}, 1789 (2011).

\bibitem{ref22} For example: K. R. Elder and M. Grant, Phys. Rev. E {\bf 70}, 051605 (2004); J. Berry, M. Grant, and K. R. Elder, Phys. Rev. E {\bf 73}, 031609 (2006); A. Adland, A. Karma, R. Spatchek, D. Buta, and M. Asta, Phys. Rev. B {\bf 87}, 024110 (2013).

\bibitem{ref23} P. Stefanovic, M. Haataja, and N. Provatas, Phys. Rev. Lett. {\bf 96}, 225504 (2006); S. Majaniemi and M. Grant, Phys. Rev. B {\bf 75}, 054301 (2007); J. Berry and M. Grant, Phys. Rev. Lett. {\bf 106}, 175702 (2011); A. Adland, Y. Xu, and A. Karma, Phys. Rev. Lett. {\bf 110}, 265504 (2013).

\bibitem{ref24} S. Majaniemi, M. Nonomura, and M. Grant, Eur. Phys. J. B {\bf 66}, 329 (2008).

\bibitem{ref25} R. L. Davidchack and B. B. Laird, J. Chem. Phys. {\bf 108}, 9452 (1998).

\bibitem{ref26} R. J. Asaro and W. A. Tiller, Metall. Trans. {\bf 3}, 1789 (1972); M. Grinfeld, Dokl. Akad. Nauk SSSR {\bf 290}, 1358 (1986) [Sov. Phys. Dokl. {\bf 31}, 831 (1986)]; Z.-F. Huang and K. R. Elder, Phys. Rev. Lett. {\bf 101}, 158701 (2008). 

\bibitem{ref27} F. Podmaniczky, G. I. T\'oth, G. Tegze, and L. Gr\'an\'asy, Metall. Mater. Trans. A {\bf 46}, 4908 (2015).    

\end{thebibliography}
\end{document}